\documentclass[12pt,preprint]{aastex}

\shorttitle{Silicate Emission Survey of Young Stars}
\shortauthors{Gaidos \& Koresko}

\begin{document}


\title{A Survey of 10-Micron Silicate Emission from Dust around Young
Sun-Like Stars \footnotemark}


\author{Eric Gaidos}
\affil{Department of Geology and Geophysics,  University of Hawaii, Honolulu, HI 96822}
\email{gaidos@hawaii.edu}

\and

\author{Christopher Koresko}
\affil{Michelson Science Center, California Institute of Technology, Pasadena, CA 91125}
\email{koresko@ipac.caltech.edu}



\begin{abstract}
We obtained low resolution (R = 100) mid-infrared (8-13 $\mu$m
wavelengths) spectra of 8 nearby young main sequence stars with the
Keck 1 telescope and Long-Wavelength Spectrometer (LWS) to search for
10 $\mu$m silicate (Si-O stretch) emission from circumstellar dust.
No stars exhibited readily apparent emission: Spectra were then
analyzed by least-squares fitting of a template based on a spectrum of
Comet Hale-Bopp.  Using this technique, we were able to constrain the
level of silicate emission to a threshold ten times below what was
previously possible from space.  We found one star, HD~17925, with a
spectrum statistically different from its calibrator and consistent
with a silicate emission peak of 7\% of the photosphere at a
wavelength of 10~$\mu$m.  Excess emission at 60 $\mu$m from this star
has already been reported.
\end{abstract}

\footnotetext[1]{Data presented herein were obtained at the W. M. Keck
Observatory, which is operated as a scientific partnership among the
California Institute of Technology, the University of California and
the National Aeronautics and Space Administration.  The Observatory
was made possible by the generous financial support of the W. M. Keck
Foundation.}




\section{Introduction \label{sec.introduction}}

One of the first, and serendipitous, discoveries of the Infrared
Astronomical Satellite ({\it IRAS}) was excess infrared emission from
circumstellar dust around main sequence stars \cite{neugebauer84}.
These disks are present well after the T Tauri phase of stellar
evolution and its associated massive accretion disk.  The lifetime of
dust grains around such stars is short compared to the ages of these
stars and their detection is thus indirect evidence for the presence
of larger source bodies that resist Poynting-Robertson drag, perhaps
analogous to asteroids or comets.  Our Solar System contains a dust
cloud (the Zodiacal Cloud) that reprocesses $f \approx 8 \times
10^{-8}$ of the Sun's radiation \cite{backman93}, re-emitting it with
a spectral energy distribution similar to that of a 260~K black-body
\cite{leinert02}.  The dust originates in collisions between asteroids
and the disintegration of short-period comets.  The early Solar System
may have contained larger numbers of minor bodies and hence elevated
levels of dust.  Ancient cratered surfaces in the inner Solar System
record intense bombardment that occured within 800 million years (Myr)
of planetary accretion \cite{strom87}.  Such impacts influenced the
evolution of inner planet surfaces, atmospheres and volatile
inventories \cite{fegley86,melosh89,chyba90}.  These impactors may
have appeared suddenly after the breakup of a larger body, or belonged
to a population with a long dynamical lifetime, perhaps in the outer
Solar System \cite{fernandez83,levison01}.  Gaidos (1999) used a model
of comet-like impactors scaled to the lunar cratering record to show
that an early Zodiacal dust cloud with $f > 3 \times 10^{-4}$ was
possible if Poynting-Robertson drag was the dominant removal
mechanism. Circumstellar dust is thus a potential probe of the
dynamics of extrasolar planetary systems and their impactor history.

The scattered or reprocessed light from residual dust disks around
young main sequence stars can be (and have been) detected, usually at
mid-infrared wavelengths where the contrast ratio with the stellar
photosphere is greatest.  {\it IRAS} detected excess 12-60 $\mu$m
emission from dust around many main-sequence A-K stars
\cite{backman93}.  Searches around solar-type (G and early K) stars
were hampered by the sensitivity limits of the {\it IRAS} telescope.
An analysis of {\it IRAS} observations of a sample of single, nearby
($d < 20$ pc) young ($< 800$ Myr) solar analogs selected on the basis
of their high coronal X-ray emission and chromospheric activity
\cite{gaidos98,gaidos00} found excess emission from only one star,
however this is now known to be the result of source confusion with a
background giant star (C. Beichmann, private communication).  {\it
IRAS} detection limits around these stars are typically $f \sim 8
\times 10^{-4}$, or $10^4$ ``Zodis''.  A slightly more sensitive
search was carried out at 25-170 $\mu$m with the Infrared Space
Observatory ({\it ISO}) (Habing {\it et al.}  2001).  That study found
that detectable ($f > 10^{-4}$) levels of dust are present around
half of stars younger than 400 Myr but only 10\% of older stars.
Three of five young ($< 800$ Myr), solar-type stars in that survey
have significant excess emission, including one star (HD 30495) in the
Gaidos (1998) sample.  This suggests that yet more circumstellar disks
will be found around young main sequence stars at lower detection
threshholds.

\begin{figure}
\plotone{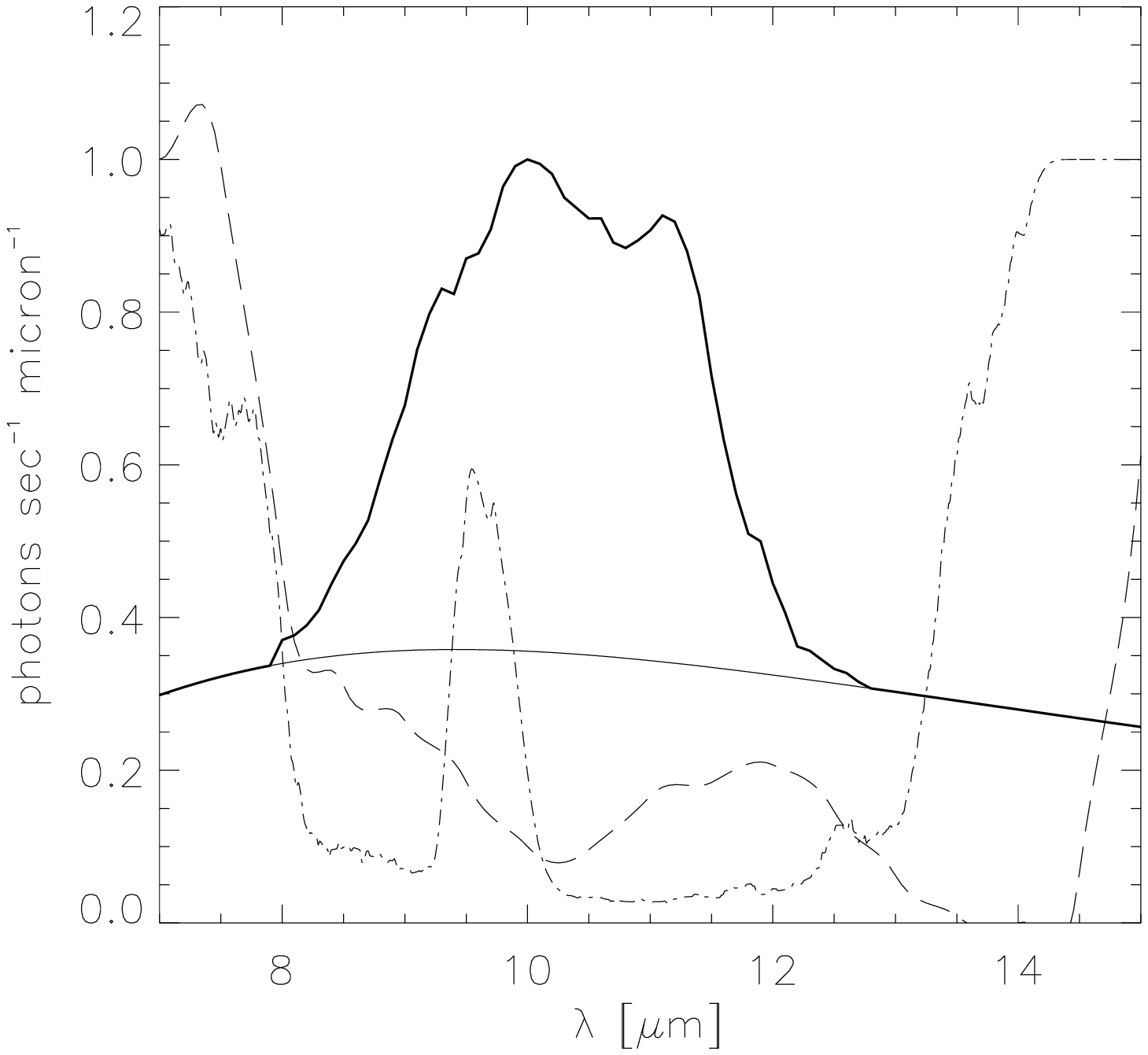}
\caption{The mid-infrared spectrum of comet Hale-Bopp, with its
pronounced silicate emission feature (Hayward et al. 2000) that
dominates the underlying continuum emission (approximated as a 390~K
blackbody).  The dash-dot line is atmospheric opacity (1 mm
precipitable water) smoothed to 0.05 $\mu$m resolution \cite{roe02},
and, for comparison with the atmospheric window, the dashed line is
one minus the response of {\it IRAS} in its 12 $\mu$m bandpass.
\label{fig.halebopp}}
\end{figure}

Ground-based detection of dust disks in the infrared is very
difficult.  At a wavelength of 10 microns, the resolution of even the
largest telescopes is equivalent to a semimajor axis of 1.3~AU at 10
pc, limiting imaging searches for dust to the nearest few bright stars
\cite{kuchner98}.  Ground-based infrared photometry is notoriously
inaccurate and the absolute mid-IR fluxes of stars cannot be precisely
predicted, rendering such an approach inutile in improving the
observational status quo.  The flux density from a hypothetical 260~K
dust cloud with $f \sim 8 \times 10^{-5}$ around a G5 star at 10 pc
is 30 mJy at 11.7 $\mu$m, representing only 6\% of the photospheric
signal at this wavelength.  A 3$\sigma$ detection would require 2\%
photometric accuracy, extremely challenging in the mid-infrared.
Alternatively, one can use mid-infrared \underline{spectroscopy} to
search for the 10~$\mu$m emission feature produced by the Si-O stretch
vibration in crystalline silicate dust.  Silicate emission features
comparable to the continuum emission have been detected from $\beta$
Pictoris and young early-type stars with circumstellar disks
\cite{skinner92,knacke93,fajardo93,waelkens96,hanner98} and resemble
those observed in comets \cite{knacke93,sitko99,hayward00}.  In
principle, spectroscopy is more precise than photometry because it is
sensitive to the wavelength dependence of the emission within the
band-pass and insenstive to the absolute total value.  The marked
features associated with crystalline silicate emission are more easily
discernable than continuum emission with a Planck or power-law
spectral energy distribution (SED) (Fig. \ref{fig.halebopp}).  In
addition, the intensity and shape of the emission feature can provide
information about the mineralogy of the dust grains \cite{jager98}.

We obtained N-band (8-13~$\mu$m) spectroscopy of several nearby young
stars using the Keck 1 telescope and the Long Wavelength Spectrometer
(LWS) to search for silicate emission from circumstellar dust.  Our
approach was to correct for the effects of the atmosphere, telescope,
and detector combination by comparing target and calibrator stars.
The ideal calibrator star would, of course, be completely dust free.
However, the presence or absence of dust around most nearby stars has
not been established at a precision comparable to that of our current
observations.  Thus, while we observed older calibration stars in the
expectation that they would be less likely to harbor significant dust
clouds, our conclusion will be limited to statements about the
\underline{relative} amount of silicate emission flux from pairs of
stars.

Most of the targets (Table \ref{table.observations}) are young solar
analogs from the updated, volume-limited version of the Gaidos (1998)
catalog of nearby young solar analogs.  These single or wide-binary
stars are all within 25 pc, have spectral types G0-K2, and coronal
X-ray luminosities suggesting ages less than 800 Myr.  Followup
spectroscopy and photometery of these objects have confirmed marked
chromospheric activity, high lithium abundance, and rapid rotation,
all indicative of stellar youth \cite{gaidos00}.  Two other stars in
the sample, HD~17925 and HD~77407 were identified as active solar-type
stars by Montes et al. (2001).  We also observed Vega ($\alpha$ Lyra)
and the calibrator A stars Altair ($\alpha$ Cygnus) and Castor
($\alpha$ Gem).  Vega is the archetypal {\it IRAS}-detected infrared
excess star and its fluxes at 25 and 60 $\mu$m exceed those predicted
for its photosphere.  Circumstellar dust may also have been resolved
at sub-millimeter wavelengths \cite{koerner01, wilner02}.  No excess
has been detected at 12 $\mu$m to the {\it IRAS} calibration accuracy
of 5\% \cite{aumann84}.  However, Ciardi et al. (2001) have reported
resolving the stellar disk of Vega with interferometry at 2.2 $\mu$m
and suggest that 3-6\% of the flux is emanating as scattered or
reprocessed starlight from circumstellar dust within 4 AU of the star.
As equilibrium black-body temperatures at those distances exceed
380~K, this material would be expected to produce a signal at
mid-infrared wavelengths as well.

\section{LWS Observations and Data Reduction \label{sec.observations}}

Observations were carried out with Keck 1 and Long Wavelength
Spectrometer (LWS) in low-resolution spectroscopic mode during two
runs; 2 September 2002, and 18-19 March 2003 (Table
\ref{table.observations}).  The gold-coated secondary has a speed of
f/25 and the LWS was mounted on the forward Cassegrain module of the
telescope.  The LWS is described in detail by Jones \& Puetter (1993).
Briefly, the LWS detector is a Boeing 128$^2$ pixel Si:As array with
75 micron pixels read out by amplifiers with a co-adder inverse gain
of 360 e$^-$ DN$^{-1}$ at 12 bit accuracy.  The low-resolution grating
was used (dispersion of 0.037 $\mu$m pixel$^{-1}$ at the detector)
with a 6-pixel (0.5 arc-second) slit and an N-wide (8.1-13.0~$\mu$m)
filter.  The detector is oriented such that rows are approximately
parallel to the direction of dispersion.

Multiple frames for each star were taken in the usual four telescope
states associated with ground-based observations in the mid-infrared,
with the chopping secondary and telescope nod set to their on-source
and off-source positions.  Typical chop amplitude and frequency were
10~arc-sec and 2~Hz.  Preprocessing of the spectral frames consisted
of subtraction of the average of the (chop off, nod on) frames in each
file from the average of the (chop on, nod on) frames to produce a
chop-subtracted nod-on frame.  Similarly, the average of the (chop
off, nod off) frames in each file was subtracted from the average of
the (chop on, nod off) frames to produce a chop-subtracted nod-off
frame.  Subtracting the nod-off frame from the nod-on frame, dividing
by the flatfield, and fixing the bad pixels produced a stellar
spectrum frame suitable for spectral extraction.  A dark current image
was constructed by averaging multiple frames with the same exposure
time and subtracted from each image.  ``Hot'' pixels were identified
in spectra of a warm stop that was placed in the beam and are removed
from further analysis.

Care was taken during the extraction of the stellar spectra to account
for the distortion of the image plane, which causes the wavelength
corresponding to each column of pixels to differ from row to row, and
the file-to-file shifts of the spectra caused by imperfect setting of
the grating and pointing of the telescope. It was important that these
effects be corrected before dividing each target spectrum by a
calibrator spectrum to avoid producing spurious spectral features,
especial in the region of deep atmospheric absorption features.  The
wavelength calibration was done using bright telluric emission
features in mid-infrared.  An airglow frame was constructed by
averaging all the frames in a file (including both chop and nod
positions), subtracting the averaged dark frame, dividing by the
flatfield frame, and fixing bad pixels.  (The sky emission is much
brighter than the stars so the airglow frame was not affected by the
presence of the stellar spectrum). The telluric lines are visible as
gently curving vertical stripes in the airglow frame. A narrow
horizontal strip across the middle of one of the airglow frames
provided a template spectrum for the airglow. This was stretched
vertically to produce an airglow template frame with the same
dimensions as the frames from which the stellar spectra were to be
extracted. The airglow lines are nearly vertical in the airglow
template frame. The airglow frame derived from each file was
numerically undistorted to fit the airglow template frame, and the
same undistortion parameters were applied to the calibrated stellar
spectrum frame for that file. This produced a set of calibrated
stellar spectrum frames with consistent wavelength scaling.

Because the alignment between the target and calibrator spectra uses
sky emission, spectral misalignment can still occur if the centroids
of the target and calibrator stars are offset within the slit.  The
slit width (6 pixels) is equivalent to 0.22 $\mu$m and thus
significant misalignment is possible.  We mitigate this effect by
performing a secondary alignment between the target and calibrator
spectra using a weighted cross-correlation in a 0.8~$\mu$m-wide window
centered on the 9.6 $\mu$m ozone feature.  (This portion of the
spectrum is not used for subsequent spectral analysis).  Typical
offsets were 0.5 pixels (0.019 $\mu$m) (Uncertainties in offset are
accounted for in \S \ref{sec.analysis})

\begin{figure}
\plotone{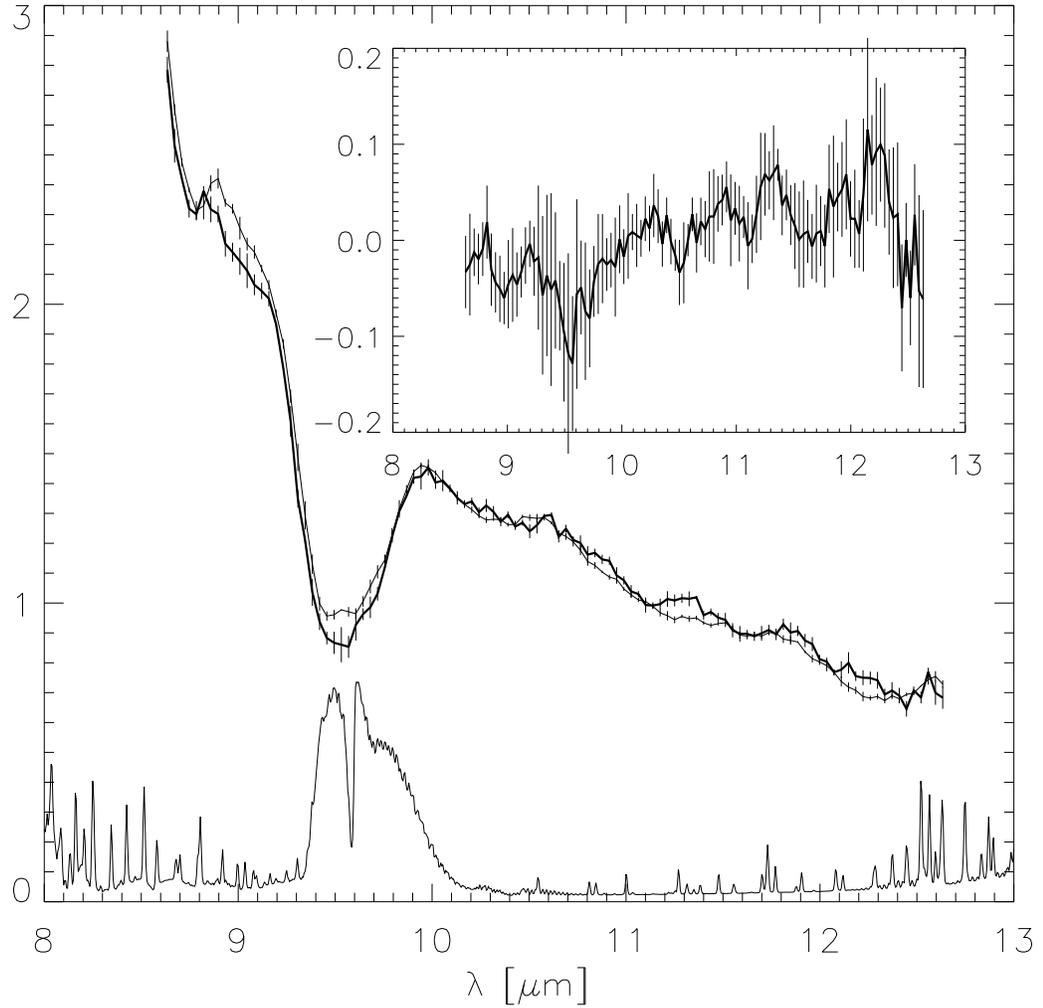}
\caption{The mid-infrared spectrum of the K0 star HD~166: The
uncalibrated, normalized spectra of HD~166 (heavy line) and the G5
calibrator star HD~20630 are plotted immediately above the atmospheric
opacity curve for the N-band \cite{roe02}.  The calibrated spectrum
(unity subtracted) of HD~166 is plotted in the inset.  The most
distinctive feature is the 9.6 $\mu$m ozone band.
\label{fig.hd166}}
\end{figure}

A spectrum was extracted from a narrow strip containing the starlight
that was projected along columns.  Each spectrum was normalized by the
average of all columns (channels).  A final spectrum was generated by
averaging over the spectra for each star.  An error spectrum was
constructing by calculating, channel by channel, the variance of the
summed spectra.  Finally, a calibrated science spectrum was generated
by taking the ratio of the normalized target and calibration spectra.
The errors in each spectral channel was calculated from the target and
calibration spectrum errors added in quadrature.  Representative
spectra are shown in Fig. \ref{fig.hd166}.  A dust-free calibrated
spectrum should be essentially flat. Differences in photosphere
temperature equivalent to an entire spectral class would induce a
slope of less than 1\% across the entire N-band.

\section{Analysis of LWS Spectroscopy \label{sec.analysis}}

In none of the spectra is silicate emission obvious, i.e., at the
level of 20\% of the photosphere or larger.  We used a matched
filter/least-squares fit technique to search for lower levels of
silicate emission from a target star.  Each calibrated spectrum $S$
(where unity has been subtracted) was compared with a template
spectrum of silicate emission $T$.  We derived the template from the
mid-infrared spectra of Comet Hale-Bopp near perihelion reported by
Hayward et al. (2000).  The comet exhibited pronounced double-peaked
silicate emission with respect to a continuum that can be described by
a 390~K blackbody (Fig. \ref{fig.halebopp}).  Crystalline olivine,
glassy olivine, and crystalline pyroxene may all contribute to the
complex structure of the emission.  The relatively narrow, intense
peak and the hot color temperature with respect to the blackbody
prediction (``superheat'') may reflect a significant contribution by
submicron particles to the mid-infrared flux \cite{hanner87}.

\begin{figure}
\plotone{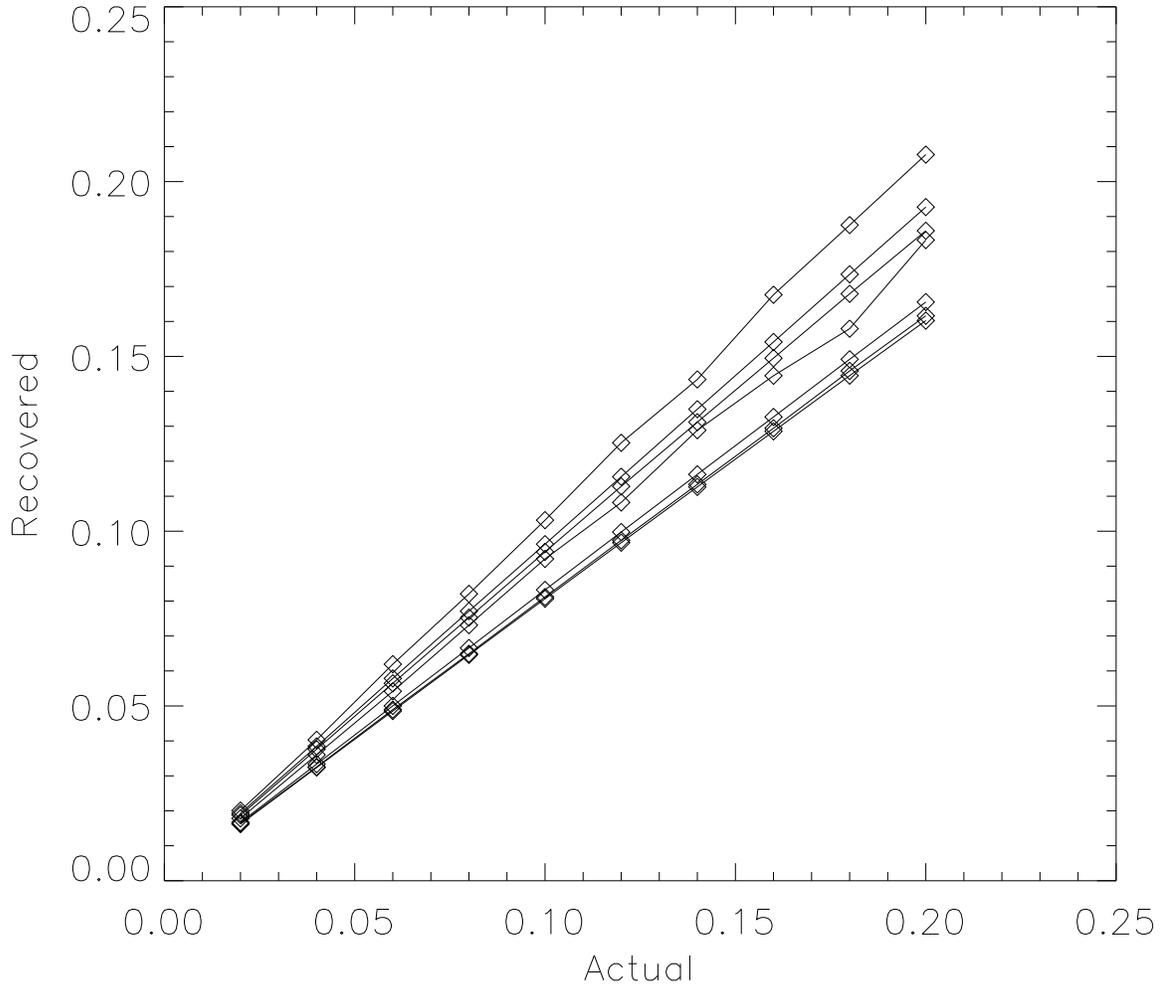}
\caption{The recovered 10 $\mu$M amplitude of a silicate emission
feature artifically added to a real stellar spectrum before
calibration by the spectrum itself, vs. the actual amplitude.  The
results from experiments with the actual spectra of seven different
stars are plotted. \label{fig.recover}}
\end{figure}

\begin{figure}
\plotone{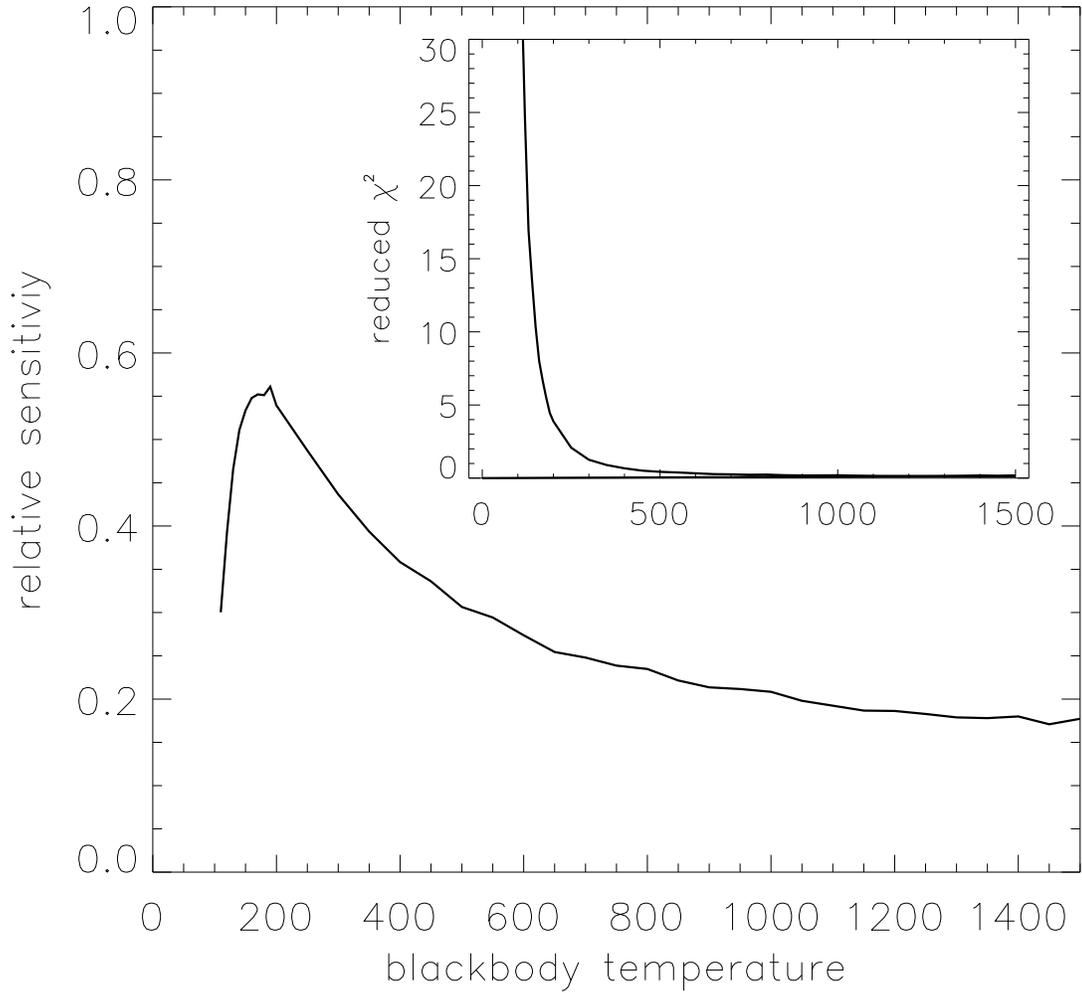}
\caption{The recovery efficiency (recovered amplitude/actual
amplitude) for blackbody spectrum added to a stellar spectrum and
detected with the silicate emission template.  Each spectrum is
normalized to 10\% of the stellar photosphere at 10 $\mu$m.  The inset
plot is the reduced $\chi^2$ \label{fig.bb}}
\end{figure}

To carry out a linear least-squares fit, we construct the template
spectrum from the comet spectrum by multiplying the latter by a small
number $\epsilon$ (here 0.01) at 10~$\mu$m.  The comet spectrum is
added to and divided by a 5600~K blackbody spectrum (representing
typical target and calibrator star), the resulting template normalized
by its average, and unity subtracted.  The least-squares fit amplitude
of the silicate emission signal at 10~$\mu$m ($A_{10}$) in a
calibrated spectrum is given by the ratios;
\begin{equation}
\frac{A_{10}}{1-0.75 A_{10}} = \epsilon\frac{\Sigma_i S_iT_i/\sigma_i^2}{\Sigma_i T_i^2/\sigma_i^2},
\end{equation} 
where the sums are over all channels.  the factor of 0.75 in the
left-hand denominator is the average of the silicate emission spectrum
relative to its 10~$\mu$m value.  A $\chi^2$ value is calculated as
follows;
\begin{equation}
\chi^2 = \Sigma_i S_i^2/\sigma_i^2 - \left[\Sigma_i S_i T_i/\sigma_i^2\right]^2\left[\Sigma_i T_i^2/\sigma_i^2\right]^{-1}
\end{equation}
The accuracy and robustness of our analysis were tested by taking one
star as both target and calibrator and artificially adding a Hale-Bopp
signal with a specified amplitude to the former.  We then compare the
original and recovered amplitudes.  Fig. \ref{fig.recover} shows that
the signal recovered from actual spectra is proportional to and close
to the actual signal.  Of course, any excess flux with an SED
different than that of the stellar spectrum will also produce a
signal.  The recovery efficiency (recovered signal/actual signal) for
excess emission with a blackbody spectrum is plotted in
Fig. \ref{fig.bb} using the star HD~20630 as both target and
calibrator.  The matched filter is most sensitive to 170-190~K
blackbody, although the $\chi^2$ will be relatively high.  Sensitivity
to hotter blackbodies is reduced as the spectrum becomes more
Raleigh-Jeans-like and resembles that of the star.  Below 150~K the
exponential shape of the SED makes any template fit meaningless and
regardless, we do not expect a significant signal in the N bandpass
from dust at these temperatures.

A detection limit was calculated for each calibrated spectrum by
repeating the least-squares fit with $10^5$ artificial spectra
generated by adding gaussian-distributed noise (with errors as
calculated above) to a perfect spectrum (unity ratio).  Systematic
errors created by the imperfect wavelength alignment between the
source and calibration spectra must also be considered.  This error
will be proportional to the local slope of the observed spectrum and
will thus be larger in regions where there are strong features, e.g.,
the ozone absorption line.  We add these errors to the artificial
target spectra by introducing a gaussian-distributed misalignment with
mean of zero and $\sigma = 0.5$ pixels before dividing the relevant
stellar spectrum by itself.  The least-squares fit amplitudes
generated from the artificial spectra are sorted and the 99.97\%
confidence level (3$\sigma$ for gaussian-distributed errors) found.

\section{Comparison to IRAS Photometry \label{sec.iras}}

Where appropriate, we estimated or put limits on the excess flux in
the {\it IRAS} 12 $\mu$m channel by fitting a theoretical SED to
visual and near-infrared photometry and comparing the predicted mid-IR
fluxes to {\it IRAS} photometry.  Visual (0.55 $\mu$m) magnitudes were
obtained from the {\it Hipparcos} catalog \cite{perryman97}.
Near-infrared magnitudes in the $J$,$H$, and $K$ passbands (1.24,
1.66, and 2.16 $\mu$m) measured by the Two Micron All Sky Survey
(2MASS) were extracted from the Infrared Science Archive (IRSA) using
the Gator query engine \cite{skrutskie01}.  Additional near-infrared
measurements were obtained from the Catalog of Infrared Observations,
Version 4.1 \cite{gezari93}.  Most stars also have 12 $\mu$m fluxes
reported in source catalogs generated from observations by the
Infrared Astronomical Satellite ({\it IRAS}) \cite{moshir90}.  The
{\it IRAS} fluxes reported in the FSC are calculated based on a
$\nu^{-1}$ SED and must be color-corrected for stars.  As in Gaidos
(1999) and others, we correct all fluxes to a 5600~K blackbody: The
color corrections for 12 and 25 $\mu$m are 0.687 and 0.714,
respectively.

Analysis was as described in Gaidos (1999).  Kurucz (1992) spectra for
solar metallicity and temperatures of 4750, 5000, 5250, 5500, 5750,
and 6000~K were linearly interpolated with 100 mesh points between
each original spectrum.  Weighted least-squares fitting to the visible
and near-infrared photometry was performed and the best-fit model used
to generate predicted fluxes at {\it ISO} and {\it IRAS} wavelengths.
Errors in these predictions were calculated by a standard Monte Carlo
approach using the errors in the photometry.  The significance of an
excess was assessed by comparing it with the independent errors in the
prediction and photometry added in quadrature.

Our LWS detections and upper limits were expressed in terms of the
equivalent flux in the 12 $\mu$m {\it IRAS} channel.  The conversion
factor (0.433) between the fractional amplitude of a silicate emission
feature at 10 $\mu$m to the fractional {\it IRAS} flux at 12 $\mu$m
(corrected to a 5600~K blackbody) was found by direct integration of
the Hale-Bopp SED over the {\it IRAS} response curve.  For example, an
infrared excess with the spectrum of Hale-Bopp and a peak amplitude of
10\% of the photosphere at 10~$\mu$m would create an excess
color-corrected flux of 4.3\% of the photosphere in the {\it IRAS}
12-micron channel.

\section{Results and Discussion}

\begin{figure}
\plotone{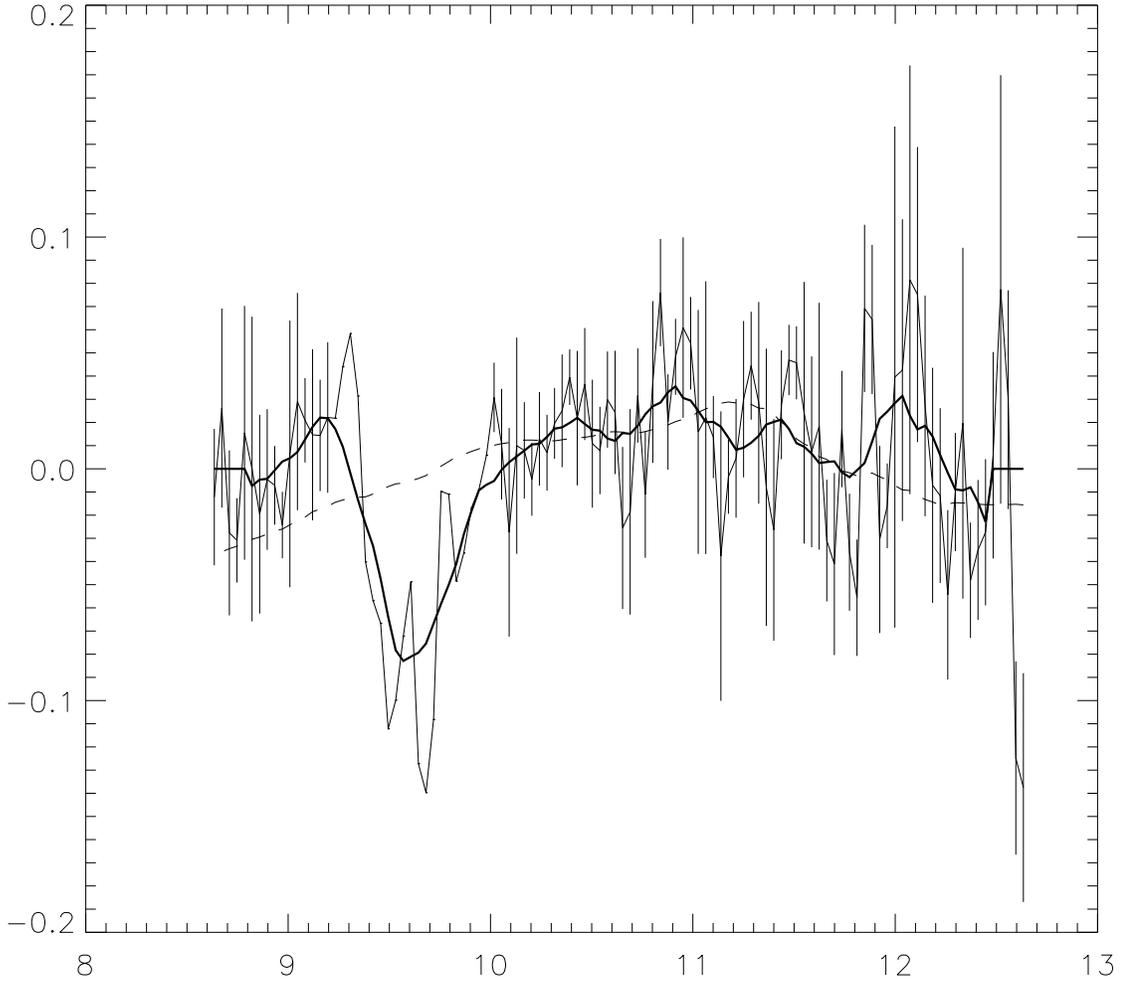}
\caption{Spectrum of HD~17925 calibrated by HD~20630 (light line).
The heavy line is a gaussian-smoothed ($\sigma$ = 5.4 pixels or R =
25) version for presentation purposes.  The dashed line is the
Hale-Bopp silicate emission template scaled to the detection
amplitude.  Error bars are absent in the region of the spectrum around
the 9.6 $\mu$m ozone line that was excluded from the
analysis. \label{fig.hd17925}}
\end{figure}

One star (HD~17925) in the observed sample has an N-band spectrum that
significantly deviates from that of the calibrator star (HD~20603).
The calibrated spectrum and the silicate emission template scaled to
the detection amplitude (7.7\% of the photosphere at 10~$\mu$m) are
shown in Fig. \ref{fig.hd17925}.  HD~17925 is a nearby (10.4 pc)
active K1 dwarf star identified as a young main-sequence star by
Montes et al. (2001).  Habing et al. (2001) report a marginally
significant (3.1$\sigma$) excess at 60~$\mu$m from {\it ISO}
observations but no excess was detected at 25~$\mu$m (Laureijs et
al. 2002).  The 2MASS JHK images of this star were saturated and we
used instead 4 measurements from the Catalog of Infrared Observations
Version 5.1 (http://ircatalog.gsfc.nasa.gov).  Photometric errors are not
given in the catalog; we assume 3\% precision.  Least-squares fitting
of an interpolated (5090~K) Kurucz model to the VJHK photometry and
comparison with {\it IRAS} and {\it ISO} photometry closely reproduces
the {\it ISO} 60 $\mu$m excess reported by Habing et al. ($78 \pm 24$
mJy).  While there is a marginal (2.2$\sigma$) excess of 50~mJy at
25~$\mu$m, there is no significant excess at 12~$\mu$m.  The {\it
IRAS} measurement is still insufficiently precise at this wavelength
($\sigma = 35$~mJy) compared to the expected excess produced by the
spectroscopic feature (23~mJy).  We note that HD~20630 (the calibrator
star) is also relatively young but does not have {\it IRAS}-detected
excess emission.

As is customary, we convert these limits into an optical depth.
Assuming the template silicate spectrum describes the actual SED of
the infrared excess the ratio of the integrated excess signal to the
photosphere within the N-band is about 0.057.  A K1 star emits about
0.13\% of its energy in the N band, therefore this detected excess
represents $f \approx 7 \times 10^{-5}$.  Of course, if this excess is
dust hotter than 300~K, it will contribute a similar amount (relative
to the photosphere) at longer wavelengths, but inclusion of this
effect increases $f$ only to $9 \times 10^{-5}$, or about 1000 times
that of our Solar System's zodiacal cloud.  For other stars we can
place $3\sigma$ limits on the amount of silicate emission of about 3\%
of the photosphere at 12 $\mu$m, a factor of ten below that of {\it
IRAS} and equivalent (per the calculations above) to $f \approx 4
\times 10^{-5}$.  Of course, these limits are only valid for a dust
SED like that of of Hale-Bopp: Our investigation is less sensitive to
continuum emission but nonetheless could detect dust at the level of
$f \approx 1 \times 10^{-4}$.  Our lack of detections suggest that
young main-sequence stars possessing warm ($\sim$200~K) circumstellar
dust with $f > 10^{-4}$ are the exception rather than the rule.  The
launch of the {\it Space Infrared Telescope} ({\it SIRTF}), currently
scheduled for 23 August 2003 and the advent of nulling imaging at the
Keck Interferometer will allow us to probe with still greater
sensitivity the dust contents of the circumstellar environments of
nearby solar-type stars.

In two of the three observations, high signal-to-noise spectra of Vega
calibrated by those of Altair produced a calibrated spectrum with a
negative slope of about -3\% across the entire N bandpass.  While the
spectrum of Castor calibrated by Altair also has a similar slope, the
spectrum of Vega calibrated by Castor appears flat.  One possible
explanation is a difference in the SED of the photosphere of Altair,
an A7 star, compared to those of the other two stars (which have
spectral types A0 and A2).  Normalization of a 9650~K blackbody (Vega)
by a 8200~K blackbody (Altair) generates a slope of only 0.8\% across
the 8-13 $\mu$m range, however stars do deviate from the ideal
black-body even in the mid-infrared.  Another, less likely,
explanation is that Altair has a circumstellar infrared excess
undetected by {\it IRAS}.  

The report of a circumstellar signal of 3-6\% of Vega at 2.2 $\mu$m
\cite{ciardi01} must be reconciled with the absence of significant
excess emission at N-band wavelengths.  A nominal isothermal blackbody
dust model predicts that for a dust albedo of 0.1 (Backman \\ \& Paresce 1993) and
a temperature of 1500~K, the excess flux at 12~$\mu$m will be $\sim$7
times that at K-band.  The {\it IRAS} limit of 5\% thus constrains the
K-band excess to $<$1\%.  Cooler dust temperatures make the constraint
even tighter.  Our detection limit of about 0.5\% for silicate
emission from Vega crudely translates into a 3\% limit on the 12
$\mu$m excess for 1500~K blackbody emission, restricting the K-band
excess to less than 0.4\%.  (The limits on cooler dust emission are
lower still).  Two possible explanations for reconciling the
observations of Ciardi et al. (2001) with the mid-infrared
observations are that the dust albedo is close to unity, implausible
for hot dust, or that the dust particles are micron-sized or smaller
and poor emitters at mid-infrared wavelengths.  Our lack of detection
of a silicate emission peak from such dust particles suggests that the
second explanation is also untenable.

{\acknowledgments

The authors are grateful to all people, and all peoples, that have
made it possible for the summit of Mauna Kea to be a premier site of
human scientific inquiry.  We thank the personnel of the William Keck
Observatory, particularly Randy Campbell, for courteous and
professional assistance in the observations.  This publication makes
use of data products from the Two Micron All Sky Survey, which is a
joint project of the University of Massachusetts and the Infrared
Processing and Analysis Center/California Institute of Technology,
funded by the National Aeronautics and Space Administration and the
National Science Foundation.  The SIMBAD database, maintained by the
Centre de Donn\'{e}es astronomiques de Strasbourg (CDS), and NASA's
Astrophysics Data Systems Bibliographic Services were used also
extensively.}

\clearpage


\clearpage

\begin{deluxetable}{llrlll}
\tabletypesize{\scriptsize}
\tablecaption{LWS OBSERVATIONS \label{table.observations}}
\tablewidth{0pt}
\tablehead{
\colhead{STAR [HD]]} & \colhead{Type$^a$} & \colhead{$f_{12}$ [Jy]$^b$}  &\colhead{Date [UT]} & \colhead{Airmass} & \colhead{Time [sec]$^c$}
}
\startdata
166 & K0 & 0.53 & 17/09/02 & 1.11-1.24 & 960\\
17925 & K1 & 0.69 & 17/09/02 & 1.31 & 480\\
20630 ($\kappa$ Ceti) & G5 & 1.40 & 17/09/02 & 1.04-1.06 & 960\\
60178 ($\alpha$ Gem)$^d$ & A2 & 7.22 & 19/09/02 & 1.08-1.10 & 360 \\
77407 & G0 & 0.20 & 19/03/03 & 1.11-1.15 & 480\\
82443 & K0 & 0.27 & 18/03/03 & 1.04-1.25 & 720\\
      &    &      & 19/03/03 & 1.01-1.02 & 600\\
97334 & G0 & 0.25 & 18/03/03 & 1.04-1.12 & 792 \\
114710 ($\beta$ Com)$^d$ & G0 & & 19/03/03 & 1.01-1.08 & 720\\
 & & & 20/03/03 & 1.04-1.21 & 720 \\
130948 & G1 & 0.46 & 18/03/03 & 1.01-1.04 & 720 \\
 & & & 20/03/03 & 1.00-1.05 & 720 \\
172167 ($\alpha$ Lyra) & A0 & 28.53 & 17/09/02 & 1.06 & 720\\
 & & & 18/03/02 & 1.21-1.35 & 360 \\ 
 & & & 19/03/03 & 1.21-1.29 & 600 \\
187642 ($\alpha$ Aql)$^d$ & A7 & 22.66 & 17/09/02 & 1.03-1.04 & 720\\
 & & & 18/09/02 & 1.28-1.35 & 216 \\
 & & & 19/09/02 & 1.28-1.42 & 600 \\
190771$^d$ & G5 & 0.40 & 17/09/02 & 1.09-1.14 & 720\\

\enddata


\tablenotetext{a}{SIMBAD}
\tablenotetext{b}{IRAS flux after 5600~K blackbody color correction}
\tablenotetext{c}{Total integration time on source w/o overheads of included frames}
\tablenotetext{d}{Designated calibrator star}

\end{deluxetable}

\clearpage

\begin{deluxetable}{llllll}
\tabletypesize{\scriptsize}
\tablecaption{SILICATE EMISSION \label{table.limits}}
\tablewidth{0pt}
\tablehead{
\colhead{TARGET} & \colhead{CALIBRATOR} & \colhead{$A_{10}^a$} & \colhead{$f_{12}$ (LWS)$^a$} & \colhead{$\chi^2$/DOF} & \colhead{$f_{12}$ (IRAS)$^a$}
}
\startdata
HD~166 & HD~20630 & [0.036] & [0.015] & --- & [0.18]\\
HD~17925 & HD~20630 & 0.070(0.10) & 0.030(0.04) & 1.1 & [\\

HD~82443 & HD~114710 & [0.058] & [0.025] & --- & [0.27]\\
HD~97334 & HD~114710 & [0.053] & [0.023] & --- & [0.23]\\
HD~130948 & HD~114710 &        &         &     & [0.18]\\
Vega & Altair & [0.011] & [0.005] & --- & 0.12$^c$ \\
Vega & Castor & [0.012] & [0.005] & --- &  \\

\enddata

\tablenotetext{a}{Fraction of photosphere level}
\tablenotetext{b}{Values in brackets are 3$\sigma$ upper limits.
Values with parantheses are uncertainties in detected fluxes}
\tablenotetext{c}{Aumann et al. 1984}


\end{deluxetable}



\end{document}